\documentclass[aps,prl]{revtex4}
\usepackage{epsfig,amsmath,graphicx}

\newcommand{\comment}[1]{}

\begin{document}

%%%%%%%%%%%%%%%%%% title page information %%%%%%%%%%%%%%%%%%
\title{Pulse shaping with birefringent crystals: a tool for quantum metrology}

\author{Guillaume Labroille$^{1,2,3}$, Olivier Pinel$^{3,4}$, Nicolas Treps$^3$ and Manuel Joffre$^{1,2}$}

\address{$^1$Laboratoire d'Optique et Biosciences, Ecole Polytechnique, Centre National de la Recherche Scientifique, 91128 Palaiseau, France}
\address{$^2$Institut National de la Sant\'e et de la Recherche M\'edicale, U696, 91128 Palaiseau, France}
\address{$^3$Laboratoire Kastler Brossel, Universit\'e Pierre et Marie Curie--Paris 6, ENS, CNRS; 4 place Jussieu, 75252 Paris, France}
\address{$^4$Centre for Quantum Computation and Communication Technology, Department of Quantum Science, The Australian National University, Canberra, ACT 0200, Australia}

%\email{} %% email address is required



\begin{abstract}
A method for time differentiation based on a Babinet-Soleil-Bravais compensator is introduced.
The complex transfer function of the device is measured using polarization spectral interferometry.
Time differentiation of both the pulse field and pulse envelope are demonstrated over a spectral width of about 100~THz with a measured overlap with the objective mode greater than 99.8\%. This pulse shaping technique is shown to be perfectly suited to time metrology at the quantum limit.
\end{abstract}

%\ocis{(320.7100) Ultrafast measurements, (320.7085) Ultrafast information processing, (320.5540) Pulse shaping.} % REPLACE WITH CORRECT OCIS CODES FOR YOUR ARTICLE

\maketitle

%%%%%%%%%%%%%%%%%%%%%%%%%%  body  %%%%%%%%%%%%%%%%%%%%%%%%%%
\section{Introduction}
%%%%%%%%%%%%%%%%%%%%%%%%%%%%%%%%%%%%%%%%%%%%%%%%%%%%%%%%

Lasers, being monochromatic or composed of ultrashort coherent train of pulses such as frequency combs, have become daily tools for high sensitivity metrology. They allow for a frequency, length or time measurement, for instance, with a sensitivity now approaching the limit imposed by the quantum nature of light. In some specific applications, such as gravitational wave detection, overcoming this limit has become an issue~\cite{GEO600}. Hence, parameter estimation with laser light at, and even beyond, the standard quantum limit has raised considerable interest in the recent years, both at the theoretical and experimental levels~\cite{Giovanetti06}.

Pioneer contributions have shown how quantum limited interferometric measurement can be performed~\cite{Caves81}, and the introduction of concepts taken from information theory has allowed for a generalization to any measurement device and quantum states: the so-called quantum Cram\'er-Rao bound~\cite{Helstrom,Braunstein94}. Proof of principle experiments are numerous and cover many different configurations. Two general trends are to be considered. Firstly, experiments performed with very dim light, where photon number and quantum states can be efficiently controlled and detected. In this regime, advantage of quantum technologies is obvious~\cite{Giovanetti06} but, given the small photon number, signal to noise ratio, and thus sensitivity, remains limited. Secondly, experiments performed with intense and coherent light. Here sensitivity scaling as $1/\sqrt{N}$, where $N$ is the number of photons, can be reached using coherent states. In this regime, getting to the standard quantum limit is already a challenge and thus quantum improvement is realistically limited to the use of gaussian states such as squeezed states~\cite{Escher}. In this regime, a lot of emphasis is put on the quality of the optical device, which should exhibit low losses and a perfect mode control. This has triggered many improvement in optical technology.

In this article, we consider parameter estimation performed with optical frequency combs, a regime of intense light for which the capacity of performing ultra-sensitive measurement is proven. Following the lines of~\cite{Delaubert08,Pinel} we know that, in this regime, to each parameter to be extracted efficiently -- being at or beyond the standard quantum limit -- only one mode of the electromagnetic field is at play. Homodyne detection whose referenced beam -- local oscillator -- has been shaped in that mode is a way to reach the standard quantum limit in parameter estimation, as shown for example in the case of space time positioning~\cite{Treps_08_prl}. Thus, sensitivity is directly linked to the ability to control precisely the optical mode, and in the present case to the quality of pulse shaping.
Although conventional pulse shaping technologies, e.g. using a programmable spatial light modulator inserted in a zero-dispersion line~\cite{Weiner_00_rsi,Chatel_10_jpb,Weiner_11_oc}, allow for versatile and accurate shaping, they suffer from spatio-temporal effects~\cite{Feurer_09_jcp} which would induce reduced visibility with the optimal mode and thus reduced sensitivity. 

In this article we propose a new and entirely optical temporal mode matching technique suitable for generating the mode required in metrology. In parameter estimation the pulse shape carrying information is known to be the derivative of the incoming train of pulses versus the parameter. For a small time delay measurement this is then simply, in frequency domain, the multiplication of the field by the frequency $\omega$. More generally, perturbation linked to the first order decomposition of the index of refraction of the medium lead to a required pulse shape mode being, in frequency domain, a multiplication by a linear function of $\omega$~\cite{Jian_12}. In this paper we propose and experimentally demonstrate the use of simple dispersive birefringent prisms to realize these pulse shapes.

The complex electric field is written as ${\cal E}(t) = {\cal A}(t) \exp(-i\omega_0 t)$, where ${\cal A}(t)$ is the pulse envelope and $\omega_0$ is the center frequency. Let us first consider two specific kinds of pulse shapes which are of interest for optimal measurements. Firstly the time derivative of the electric field, which is associated to time delay in vacuum: 
\begin{equation}
{\cal E}_1(t) = T_1 \frac{d{\cal E}}{dt},
\end{equation}
secondly the pulse corresponding to the time derivative of the envelope, which is associated to time of flight measurement:
\begin{equation}
{\cal E}_2(t) = T_2 \frac{d{\cal A}(t)}{dt} \exp(-i\omega_0 t).
\end{equation}
The time constants $T_1$ and $T_2$ have been introduced for sake of dimensional homogeneity.
In frequency domain, we thus have to fulfill the relations
\begin{equation}
{\cal E}_1(\omega) = - i\omega T_1 {\cal E}(\omega)
\end{equation}
and
\begin{equation}
{\cal E}_2(\omega) = - i(\omega-\omega_0) T_2 {\cal E}(\omega).
\end{equation}
Such pulses can be generated using linear filters with transfer functions equal respectively to $R_1(\omega) = - i \omega T_1$ and $R_2(\omega) = - i(\omega-\omega_0)T_2$. Note however that, in the case of passive linear filters, the transfer function must have a magnitude smaller than or equal to 1, which will require a compromise between accuracy and efficiency. Especially in the latter case where the transfer function increases in the wings of the spectrum, the transfer function will have to be clipped in order to maintain a reasonable overall efficiency of the linear filter.

All-optical implementation of such transfer functions in devices known as time differentiators has been reported earlier~\cite{Muriel_08_ol,Slavik_07_ol,Azana_07_ol} using fiber-based methods and is useful for information processing or pulse characterization.
In this article, we show that the use of birefringent materials can extend the achievable spectral width and thus constitutes an interesting method for the delivery of ultrashort shaped time-derivated pulses.

%%%%%%%%%%%%%%%%%%%%%%%%%%%%%%%%%%%%%%%%%%%%%%%%%%%%%%%%
\section{Time differentiation using birefringence}
\label{SecPrinciple}
%%%%%%%%%%%%%%%%%%%%%%%%%%%%%%%%%%%%%%%%%%%%%%%%%%%%%%%%

The basic principle for producing the time derivative of an ultrashort pulse ${\cal E}(t)$ relies on the destructive interference between two replica of the incident pulse separated by a time delay $\tau$ much smaller than the optical cycle:
\begin{equation}
\frac{1}{2}{\cal E}(t+\tau/2) - \frac{1}{2}{\cal E}(t-\tau/2) \approx \frac{\tau}{2} \frac{d{\cal E}}{dt}.
\end{equation}
Note that the $1/2$ factor is introduced to account for the fact that the energy transmission of each arm of the interferometer is $1/4$, when taking into account the beam splitter and the beam recombiner.
This equation yields the desired time derivative ${\cal E}_1(t)$, with $T_1 = \tau/2$.
The time derivative of the pulse envelope can be similarly obtained by choosing $\tau$ equal to a multiple of the optical cycle, {\it i.e.} $\tau  = 2 n \pi /\omega_0$, where $n$ is an integer number. Provided that $\tau$ is much smaller than the pulse duration, we obtain
\begin{equation}
\frac{1}{2}{\cal E}(t+\tau/2) - \frac{1}{2}{\cal E}(t-\tau/2) \approx \frac{\tau}{2} \frac{d{\cal A}}{dt} \exp(- i \omega_0 t).
\end{equation}
This yields the desired field ${\cal E}_2(t)$, with $T_2 = \tau/2$.

\begin{figure}[htbp]
\centering\includegraphics[width=12cm]{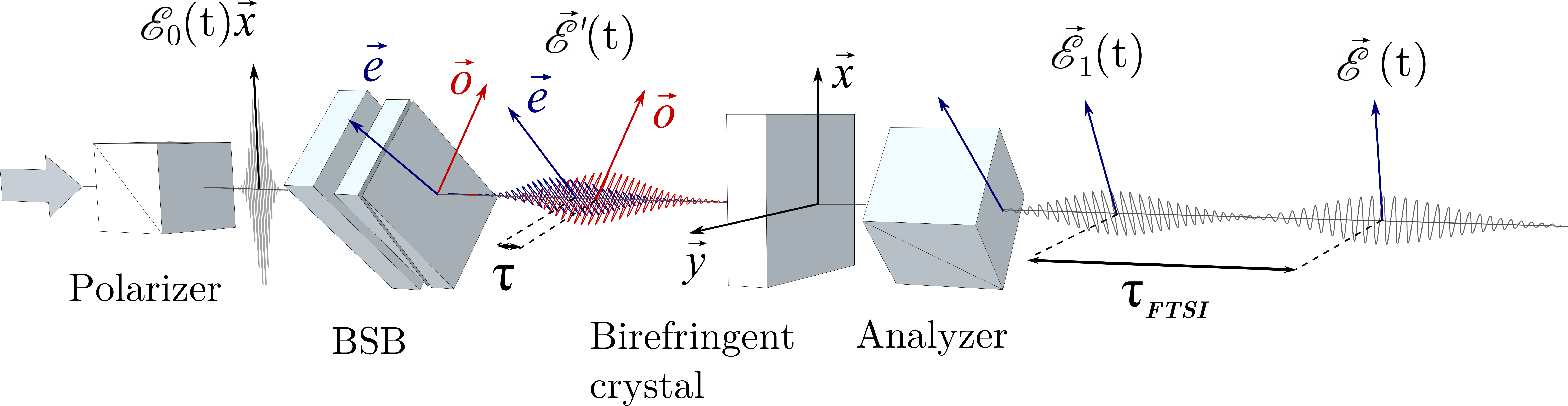}
\caption{Experimental setup. A linearly polarized pulse ${\cal E}_0(t) \vec x$ is sent through a Babinet-Soleil-Bravais compensator (BSB) rotated by 45$^\circ$ to produce a pair of orthogonally polarized pulses $\vec {\cal E}'(t)$ with a relative time delay $\tau$. For phase measurements, a large time delay $\tau_{FTSI}$ is introduced between the two pulses of interest, polarized along $\vec x$ and $\vec y$, with a birefringent crystal aligned on $\vec x$. An analyzer at 45$^\circ$ allows recording the interference spectrum between the two pulses.}
%\label{FigPrinciple}
\label{FigSetup}
\end{figure}

In earlier work, the generation of two time-delayed replicas has been demonstrated e.g. with an interferometer.
It is well known that the generation of two time-delayed pulses can also be achieved by use of birefringence in a uniaxial crystal, as has been applied for example in the case of polarization spectral interferometry~\cite{Dorrer_98_josab}.
In order to avoid using a birefringent crystal of too small thickness and to be able to easily adapt its actual length, we choose a Babinet-Soleil-Bravais (BSB) compensator -- a device consisting of two birefringent prisms whose extraordinary axes are perpendicular -- but the theory discussed below is independent of that choice. A related approach has been recently demonstrated for the production of time-delayed pulses in multidimensional spectroscopy~\cite{Cerullo_12_ol}, although we are interested here in much smaller values of the time delay.

Let us model the propagation of light through a birefringent material, whose ordinary ($\vec o$) and extraordinary ($\vec e$) axes make a $45^\circ$ angle with respect to the vertical. We consider an initial pulse polarized along the vertical axis ($\vec x$), $\vec {\cal E}_0(t) = {\cal E}_0(t) \vec x =  {\cal E}_0(t)/\sqrt{2}\ \vec o +{\cal E}_0(t) /\sqrt{2}\ \vec e$. After propagation through a birefringent material of thickness $L$, the transmitted electric field then reads
\begin{equation}
\vec {\cal E}'(\omega) =
\frac{{\cal E}_0(\omega)}{\sqrt{2}} \exp(i k_o(\omega) L) \vec o 
+
\frac{{\cal E}_0(\omega)}{\sqrt{2}} \exp(i k_e(\omega) L) \vec e
\end{equation}
or, in the initial reference frame
\begin{equation}
\vec {\cal E}'(\omega) =
{\cal E}_0(\omega) e^{i \varphi(\omega) }
\left(
\cos\left(\frac{\delta k(\omega) L}{2}\right) \vec x
+
i \sin\left(\frac{\delta k(\omega) L}{2}\right) \vec y
\right)
\label{EqCosSin}
\end{equation}
where $\delta k(\omega) = k_e(\omega) - k_o(\omega)$ and $\varphi(\omega) = (k_e(\omega)+k_o(\omega))L/2$.
Assuming that the thickness $L$ is small enough so that $\delta k(\omega) L$ is always much smaller than $\pi/2$, we can expand the above expression up to first order
\begin{equation}
\vec {\cal E}'(\omega) \approx
{\cal E}_0(\omega) e^{i \varphi(\omega) }
\left(
\vec x
+
i \frac{\delta k(\omega) L}{2} \vec y
\right)
\end{equation}
Finally, we may expand the wavevector difference as a function of frequency around the center frequency, $\delta k(\omega) = \delta k(\omega_0) + (\omega-\omega_0) \delta k'(\omega_0) = (\omega-\omega_1)\delta k'(\omega_0)$, where
\begin{equation}
\omega_1 =  \omega_0 - \frac{\delta k(\omega_0)}{\delta k'(\omega_0)} = \frac{\delta n_g(\omega_0)-\delta n(\omega_0) }{\delta n_g(\omega_0)} \omega_0
\end{equation}
where $\delta n(\omega)$ and $\delta n_g(\omega)$ are respectively the differences in phase index and group index between the extraordinary and ordinary axes.

In the context of parameter estimation, we recall that pulse shaping should occur on the local oscillator, {\it i.e.} the reference field that will interfere with the signal field in order to extract information. Thus, it is the relative pulse shape of the local oscillator versus the signal which is important. In the present case, we can decide that the signal field corresponds to the field polarized along the $x$ axis and the shaped local oscillator field corresponds to the field polarized along the $-y$ axis, {\it i.e.} $\vec{\cal E}' = {\cal E} \vec x - {\cal E}_1 \vec y$. We then obtain a transfer function equal to $R_1(\omega) = -  i (\omega - \omega_1) \delta k'(\omega_0) L/2$.
For a non dispersive material, the phase index and group index would be identical and the device would produce a pure time shift, as desired.
As expected, the value of the time shift is $\tau = 2 T_1 = \delta k'(\omega_0) L$, {\it i.e} the difference in group delay between the two polarizations.

For an actual material, the difference between the phase and group index will slightly change the response function. In the case of quartz, at a center wavelength of 800~nm, we have $\delta n \approx 8.9\times 10^{-3}$ and $\delta n_g \approx 9.5 \times 10^{-3}$ so that $\omega_1 \approx  0.063 \omega_0$, corresponding to a frequency of 24~THz only. This is reasonably close to zero so that the achieved transfer function can be considered satisfactory.

In order to generate the time derivative of the pulse envelope, we can choose a thickness $L$ such that the optical path difference is a multiple of the center wavelength: $\delta k(\omega_0) L/2 = n \pi$, where $n$ is an integer number.
Expanding $\delta k(\omega)$ around the center frequency at first order, {\it i.e.} $\delta k(\omega) = \delta k(\omega_0) + (\omega-\omega_0) \delta k'(\omega_0)$, eq.~\ref{EqCosSin} yields
\begin{equation}
\vec {\cal E}'(\omega) \approx
(-1)^n 
{\cal E}_0(\omega) e^{i \varphi(\omega) }
\left(
\vec x
+
i (\omega-\omega_0) \delta k'(\omega_0) \frac{L}{2} \vec y
\right)
\end{equation}
Again, choosing ${\cal E}$ along the $x$ axis and ${\cal E}_2$ along the $-y$ axis yields the transfer function $R(\omega) = - i (\omega-\omega_0) T_2$, with $T_2 = \delta k'(\omega_0) L/2$, which corresponds to a time shift $\tau = 2 T_2$ equal to the group delay difference between the two polarizations.
Greater values of the order $n$ are associated with greater thicknesses $L$ and thus a better transmission, at the cost of accuracy since the expansion performed above will become less appropriate.
A greater accuracy can also be achieved by use of the half-integer order value $n=1/2$. This corresponds to a case where the phase shift is close to $\pi/2$, so that the cosine in eq.~\ref{EqCosSin} can be linearized whereas the sine can be approximated with 1.
At first order, using the relation $\cos(\pi/2+\alpha) \approx -\alpha$, we thus obtain
\begin{equation}
\vec {\cal E}'(\omega) =
i {\cal E}_0(\omega) e^{i \varphi(\omega) }
\left(
 i (\omega-\omega_0) \delta k'(\omega_0) \frac{L}{2} \vec x
+
\vec y
\right)
\end{equation}
Choosing ${\cal E}$ now along the $-y$ axis and ${\cal E}_2$ along the $x$ axis allows getting the desired transfer function, with a greater accuracy at the cost of a smaller efficiency.

%%%%%%%%%%%%%%%%%%%%%%%%%%%%%%%%%%%%%%%%%%%%%%%%%%%%%%%%
\section{Experimental setup}
%%%%%%%%%%%%%%%%%%%%%%%%%%%%%%%%%%%%%%%%%%%%%%%%%%%%%%%%

The experimental setup is shown in Fig.~\ref{FigSetup}. The laser beam, produced by a Ti:Sapphire oscillator (about 10 fs Fourier-transform-limited pulses, 75 MHz repetition rate, Synergy PRO, Femtolasers) is linearly polarized using a Glan-Thompson polarizer (GTH10M, Thorlabs) specified for an extinction ratio of 10$^{-5}$. A quartz BSB compensator (Fichou), rotated by 45$^\circ$, is used to control the effective length $L$ of birefringent material. An analyzer aligned along the horizontal or vertical direction then allows the measurement of the appropriate amplitude spectra,
using an integrating sphere (FOIS-1, Ocean Optics) and a spectrometer (Acton SP2300, Princeton Instruments).

The phase of the complex transfer function corresponds to the relative phase between the two fields, which happen to be polarized along perpendicular directions.
This phase can thus be measured in a straightforward way by use of polarization spectral interferometry~\cite{Dorrer_98_josab}, which consists of inserting a birefringent crystal in order to introduce a time delay between the two pulses.
The analyzer, now aligned at 45$^\circ$, allows recording the interference spectrum between the two pulses, thus yielding the relative spectral phase by use of Fourier-Transform Spectral Interferometry (FTSI)~\cite{Joffre_95_josab,Joffre_00_josab}.
As shown in Fig.~\ref{FigSetup}, we applied this method by inserting a YVO$_4$ crystal whose neutral axis is vertical.
Taking the difference between the spectral phase obtained with the BSB inserted in the path and the spectral phase obtained when the BSB is not present allows subtracting the difference in spectral phase between the two axes of the YVO$_4$ crystal.

%%%%%%%%%%%%%%%%%%%%%%%%%%%%%%%%%%%%%%%%%%%%%%%%%%%%%%%%
\section{Time differentiation of the pulse electric field}
%%%%%%%%%%%%%%%%%%%%%%%%%%%%%%%%%%%%%%%%%%%%%%%%%%%%%%%%

\begin{figure}[htbp]
\centering\includegraphics[width=12cm]{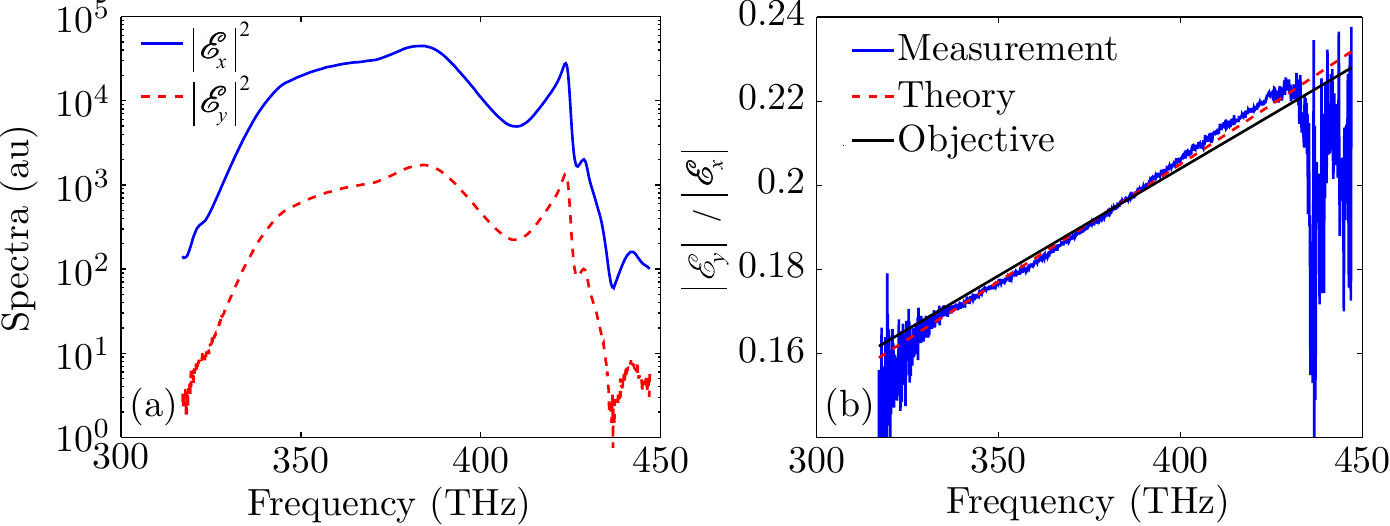}
\caption{Time differentiation of the pulse field. (a) Power spectrum measured along the $x$ (blue solid line) and $y$ polarizations (red dashed line), corresponding respectively to unshaped and shaped pulses. (b) Measured ratio of the field amplitudes (blue solid line), as compared to the objective (back solid line) and calculation (red dashed line).}
\label{FigSpectra1}
\end{figure}

In the first set of experiments, our objective was to produce the time derivative ${\cal E}_1(t)$ of the electric field ${\cal E}(t)$.
We set the BSB compensator to a differential quartz thickness of $\delta L = 5.4\;\mu{\rm m}$, which corresponds to a time delay $\tau = 0.17\;{\rm fs}$.
We then measured the spectra along the $x$ and $y$ polarizations, corresponding respectively to $|{\cal E}(\omega)|^2$ and $|{\cal E}_1(\omega)|^2$, as shown in Fig.~\ref{FigSpectra1}(a).
As predicted by theory, the overall spectral amplitude of the shaped pulse is significantly smaller, corresponding to about 3.6\% as compared to the other polarization.
Fig.~\ref{FigSpectra1}(b) shows the square root of the ratios of the two spectra, corresponding to $|{\cal E}_1(\omega)| / |{\cal E}(\omega)| = |R_1(\omega)|$. Here again, as expected, the result is very close to the objective $\omega T_1$.

\begin{figure}[htbp]
\centering\includegraphics[width=12cm]{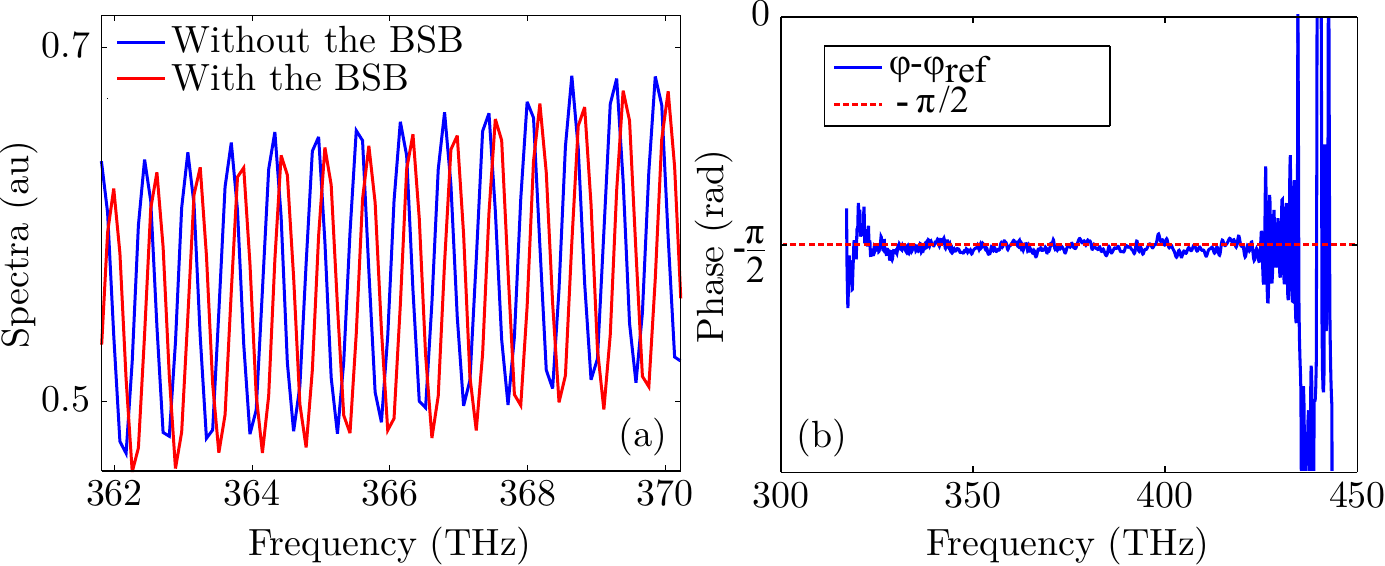}
\caption{(a) Fraction of the interference spectrum between the unshaped and shaped pulses, measured with and without the BSB compensator. (b) Spectral phase retrieved using FTSI.}
\label{FigPhase1}
\end{figure}

We then proceeded in measuring the phase of the complex transfer function by inserting the birefringent YVO$_4$ crystal as explained in the previous section. The spectral fringes obtained without and with the BSB are shown in Fig.~\ref{FigPhase1}(a) on a restricted spectral range. It is clear that the two sets of fringes are in quadrature, as expected for the purely imaginary transfer function $R_1(\omega) = - i \omega T_1$. Fig.~\ref{FigPhase1}(b) shows the spectral phase produced after applying the FTSI processing, after subtracting the reference phase obtained in absence of the BSB so that dispersion in the YVO$_4$ crystal cancels out. The result is indeed frequency independent and takes the expected value of $-\pi/2$.

The experimental results shown in Fig.~\ref{FigSpectra1}(b) and Fig.~\ref{FigPhase1}(b) thus directly demonstrate that the transfer function is proportional to the desired value of $-i \omega T_1$.
 A complete calculation based on the Sellmeier relations for quartz taking into account the difference between the phase index and the group index predicts more than $99.99\%$ overlap with the objective mode, while we measured $99.83\%$.

%%%%%%%%%%%%%%%%%%%%%%%%%%%%%%%%%%%%%%%%%%%%%%%%%%%%%%%%
\section{Time differentiation of the pulse envelope}
%%%%%%%%%%%%%%%%%%%%%%%%%%%%%%%%%%%%%%%%%%%%%%%%%%%%%%%%

\begin{figure}[htbp]
\centering\includegraphics[width=12cm]{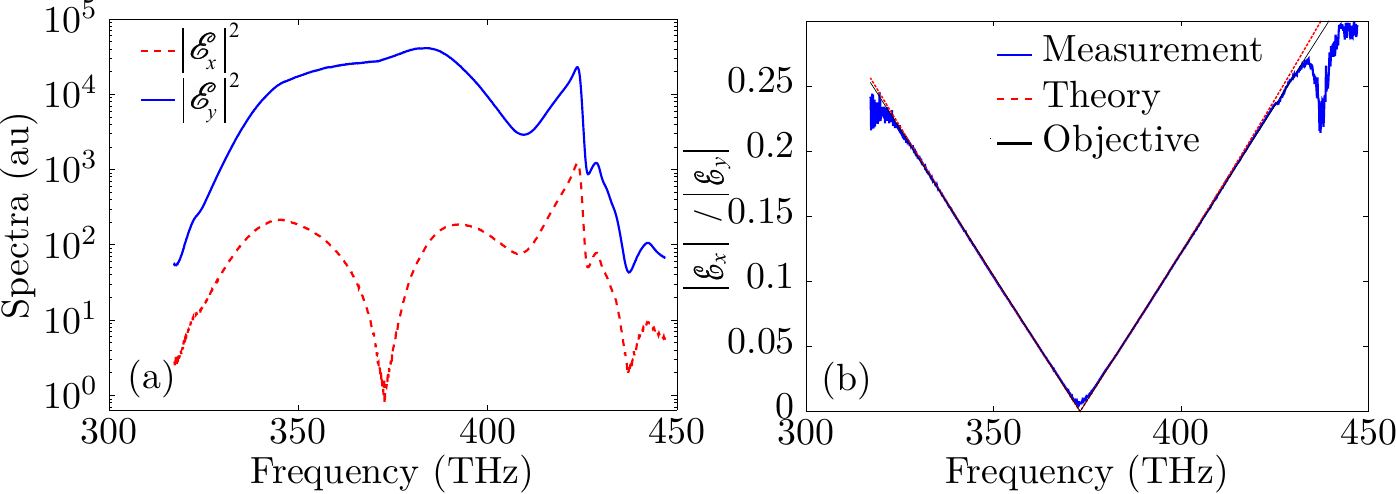}
\caption{Time differentiation of the pulse envelope. (a) Power spectrum measured along the $y$ (blue solid line) and $x$ polarizations (red dashed line), corresponding respectively to unshaped and shaped pulses. (b) Measured ratio of the field amplitudes (blue solid line), as compared to the objective (back solid line) and calculation (red dashed line).}
\label{FigSpectra2}
\end{figure}

In order to produce the time derivative of the pulse envelope, we proceed as explained in section~\ref{SecPrinciple} and increase the thickness $\delta L$ in the BSB until we observe a zero transmission in the spectrum measured along the $x$ axis, corresponding to an order $n = 1/2$.
We then adjust the point of zero transmission to the center frequency $\omega_0$, which corresponds in quartz to a differential thickness of $\delta L = 45\;\mu{\rm m}$.
Fig.~\ref{FigSpectra2}(a) shows the measured spectra with the analyzer along the $y$ axis ($|{\cal E}(\omega)|^2$) and along the $x$ axis ($|{\cal E}_2(\omega)|^2$). Note that the two axes have been exchanged in order to account for the half-integer value of the order $n$.
Fig.~\ref{FigSpectra2}(b) shows the amplitude ratio which is clearly proportional to $|\omega-\omega_0|$.

\begin{figure}[htbp]
\centering\includegraphics[width=8cm]{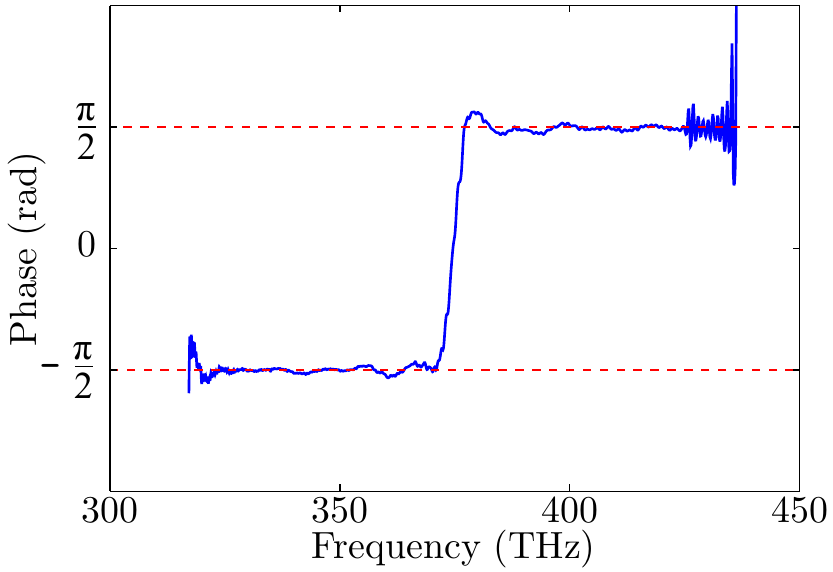}
\caption{Spectral phase retrieved using polarization spectral interferometry, evidencing the $\pi$ phase jump when the term $(\omega-\omega_0)$ changes sign.}
\label{FigPhase2}
\end{figure}

Fig.~\ref{FigPhase2} shows a measurement of the spectral phase of the transfer function, obtained using the same protocol as in the previous section. It exhibits the $\pi$ phase shift expected for the change of sign of the function $\omega-\omega_0$ when crossing the center frequency.
However, the desired transfer function $R_2(\omega) = -i(\omega-\omega_0) T_2$ would yield a $-\pi$ phase jump instead of the observed $+\pi$ phase jump. This discrepancy is of no importance and merely relates to a different sign convention on the horizontal axis, which could occur for example after a reflection on a mirror, or a $\pi/2$ rotation of the analyzer.
Another issue is the small overshoot in the phase jump, which is however not associated with a great error in the complex transfer function since the amplitude is close to zero in this spectral range.  Here the complete calculation predicts $99.99\%$ overlap with the objective mode, while we measure $99.86\%$.

%%%%%%%%%%%%%%%%%%%%%%%%%%%%%%%%%%%%%%%%%%%%%%%%%%%%%%%%
\section{Conclusion}
%%%%%%%%%%%%%%%%%%%%%%%%%%%%%%%%%%%%%%%%%%%%%%%%%%%%%%%%

We have demonstrated time differentiation of the pulse electric field and of the pulse envelope by use of a Babinet-Soleil-Bravais compensator. The amplitude and phase of the complex transfer function has been measured over a spectral bandwidth of about 100~THz by use of polarization Fourier-Transform Spectral Interferometry and has been shown to be in very good agreement with the target value. An even better result could be obtained by combining two BSB compensators made of different materials, such as quartz and KDP for example. This approach, well established for designing achromatic waveplates, would thus allow shifting the value of $\omega_1$ either to zero or to the carrier frequency in order to provide an even more exact transfer function.

The time differentiator demonstrated in this article fits perfectly the needs of high sensitivity space time positioning following the scheme of~\cite{Treps_08_prl}, being completely passive and free of spatio-temporal coupling.

%%%%%%%%%%%%%%%%%%%%%%%%%%%%%%%%%%%%%%%%%%%%%%%%%%%%%%%%
\section*{Acknowledgment}
%%%%%%%%%%%%%%%%%%%%%%%%%%%%%%%%%%%%%%%%%%%%%%%%%%%%%%%%

This work was supported by Agence Nationale de la Recherche project QUALITIME (ANR-09-BLAN-0119), European ERC starting grant program Frecquam.

% and by the Australian Research Council Centre of Excellence for Quantum Computation and Communication Technology, project number CE110001027.

%%%%%%%%%%%%%%%%%%%%%%% References %%%%%%%%%%%%%%%%%%%%%%%%%

\end{document}